\documentclass[sigchi]{acmart}
\AtBeginDocument{%
  }

\setcopyright{acmlicensed}
\copyrightyear{2025}
\acmYear{2025}
\acmDOI{XXXXXXX.XXXXXXX}
\acmConference[CHI '25]{Human Factors in Computing Systems}{April 26 -- May 01,
  2025}{Yokohama, Japan}
\acmISBN{978-1-4503-XXXX-X/2025/05}

\begin{document}

\title{Utilizing Virtual Reality for Wildfire Evacuation Training}


\author{Alison Crosby}
\affiliation{%
  \institution{University of California, Santa Cruz}
  \city{Santa Cruz}
  \state{CA}
  \country{USA}}
\email{arcrosby@ucsc.edu}

\author{MJ Johns}
\affiliation{%
  \institution{University of California, Santa Cruz}
  \city{Santa Cruz}
  \state{CA}
  \country{USA}}
\email{mljohns@ucsc.edu}

\author{Katheine Isbister}
\affiliation{%
  \institution{University of California, Santa Cruz}
  \city{Santa Cruz}
  \state{CA}
  \country{USA}}
\email{katherine.isbister@ucsc.edu}

\author{Sri Kurniawan}
\affiliation{%
  \institution{University of California, Santa Cruz}
  \city{Santa Cruz}
  \state{CA}
  \country{USA}}
\email{skurnia@ucsc.edu}

\renewcommand{\shortauthors}{Crosby et al.}

\begin{abstract}
  The risk of loss of lives and property damage has increased all around the world in recent years as wildfire seasons have become longer and fires have become larger. Knowing how to prepare and evacuate safely is critical, yet it may be daunting for those who have never experienced a wildfire threat before. This paper considers the potential for utilizing virtual reality (VR) technology to prepare people for an evacuation scenario. We discuss the unique affordances of VR for this type of work, as well as the initial steps in creating a training simulation. We also explore the next steps for what a tool like this may mean for the future of evacuation preparedness training.
\end{abstract}


\begin{teaserfigure}
  \includegraphics[width=\textwidth]{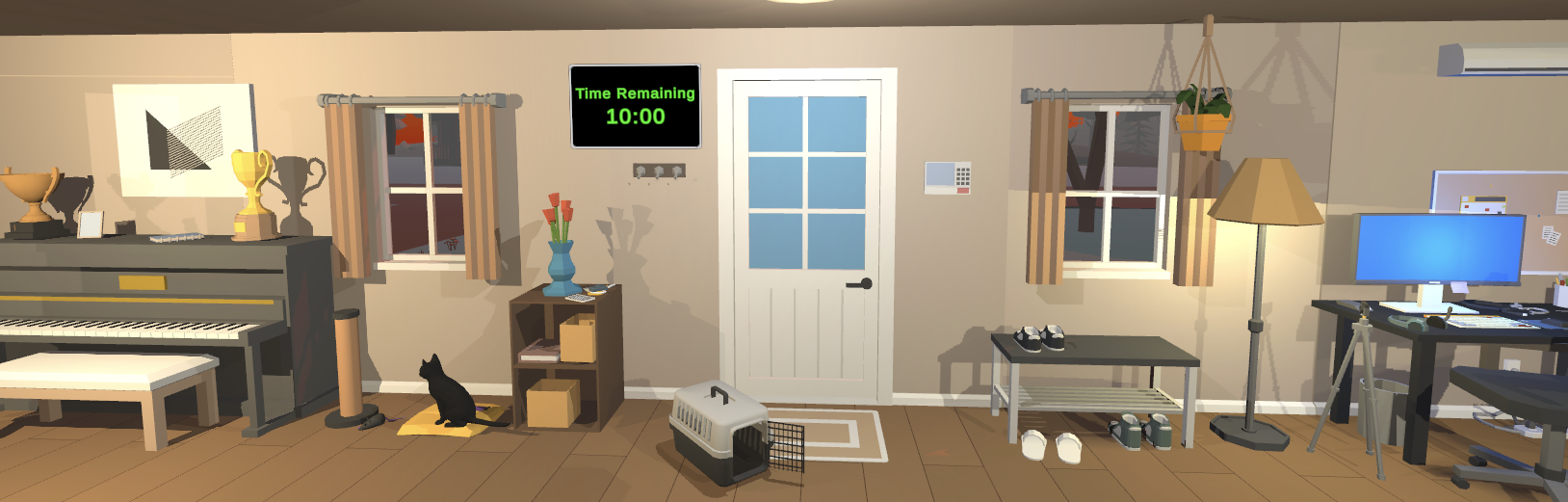}
  \caption{Scene from the virtual reality environment.}
  \Description{The player looks at the cat with the time remaining to evacuate on the background of the wall. Outside the window, a red glow from the fire can be seen.}
  \label{fig:teaser}
\end{teaserfigure}

\received{01 March 2025}

\maketitle

\section{Introduction}
Virtual reality (VR) has proven to be a powerful tool for training professionals in various fields. The effectiveness of VR stems from its ability to recreate stressful scenarios that allow users to practice decision-making skills with consequences in a safe, controlled environment that enables key training features like time compression and consistency. The benefits of VR rely on its ability to create “tasks [that] require a combination of presenting visualizations while being able to interact and manipulate objects within the environment” \cite{abich2021review}.

Several studies have shown the effectiveness of utilizing VR for training medical professionals like surgeons practicing laparoscopic surgeries \cite{larsen2009effect} and emergency responders practicing mass causality events \cite{lochmannova2022use}. VR is also commonly used in training for high-risk professions like mine workers \cite{grabowski2015virtual} and with police officers \cite{zechner2023enhancing}.
In addition, significant research has involved VR in training people for disasters. For example, quick clay landslides \cite{alene2024quickaware}, evacuating from burning buildings \cite{shaw2019heat} and homes \cite{bourhim2020efficacy}, as well as practicing earthquake safety maneuvers \cite{li2017earthquake}. 
VR is a beneficial tool for these high-risk, high-stress scenarios because of its immersive and engaging nature that brings a sense of realism to training while keeping participants safe from dangerous situations. Additionally, creating virtual training experiences is often less expensive than creating realistic but safe training environments in the real world. 

Due to the climate crisis, natural hazards like hurricanes and wildfires are expected to pose a greater and greater risk to human life, resulting in loss of life and costly economic damages \cite{balaguru2023increased, kurvits2022spreading}. Specifically looking at wildfires, in 2022, the United Nations Environment Programme (UNEP)
released the report \textit{Spreading like Wildfire: The Rising Threat of Extraordinary Landscape Fires}\footnote{United Nations Environment Programme (2022). Spreading like Wildfire – The Rising Threat of Extraordinary Landscape Fires. A UNEP Rapid Response Assessment. Nairobi.} detailing the increasing threat and prominence of wildfires in areas not previously affected. While mitigation and management are necessary for wildfire threat prevention, the expected increasing threat is already set. Therefore, ensuring people are prepared to act accordingly when threatened by wildfires is of the utmost necessity, especially for those previously unaffected. Just as VR has become a tool for training professionals for high-risk, high-stress scenarios, it can also become a tool for the general public to engage in evacuation scenarios to help them be more prepared.

In addition, the accessibility of VR is more widespread than ever before. VR head-mounted display (HMD) devices like the Quest\footnote{\url{https://www.meta.com/quest/}} and technology utilizing phones like Google Cardboard\footnote{\url{https://arvr.google.com/cardboard/}} allow for these systems to be relatively portable and affordable. This means that training simulations can be utilized and exhibited in conferences and community spaces to raise awareness of the magnitude and seriousness of natural hazard training. This increases its potential to be an effective training tool for firefighters and emergency response organizations to engage with communities alongside their other preparedness/ awareness initiatives. Not everyone can retain information from virtual web training, just as not everyone can use VR. More training options are continuously needed because of the increase in wildfire risks and impacts on human life.  VR can also potentially reach a younger audience, like college students, whom more traditional awareness approaches may not engage. 

\section{Designing for Wildfire Evacuations}


Current methods for wildfire preparedness training involve paper pamphlets, informational websites and videos, and in-class instruction \cite{hsu2013state}.
As noted, risks for natural hazards like wildfires are notably increasing in areas not previously at risk. This means it is more important than ever that information concerning wildfire preparedness training and evacuation techniques be disseminated through all forms of media. 

There are several steps that people can take to be more safe and knowledgeable about wildfire risk. For instance, fire protection agencies recommend being prepared by ensuring a house or building maintains a defensible space - meaning it is cleared of fallen debris, the gutters are cleaned out, and easily ignitable plants are not near the house. Additional steps include hardening one's home by building or remodeling with fire-resistant materials, like a metal roof. Beyond maintaining a safe home, people should also be prepared by creating an evacuation plan with everyone in a household, having an emergency supply kit on hand, and having a go bag prepared. A go bag is for anyone living in areas susceptible to natural hazards, containing things like water, cans of food, copies of important documents, spare medication, phone chargers, etcetera. These bags are important during a disaster event, as one can just \textit{grab the bag} and \textit{go}.

Packing and preparing ahead of an event like a wildfire evacuation is extremely important as prepared individuals can reduce the risk of injuries, death, and damages \cite{fazeli2024role}. If there is a delay in pre-evacuation behavior, then additional strain is placed on emergency responders as civilians can become trapped in dangerous zones \cite{zhao2009post}. 
We believe the affordances of VR can be utilized to positively impact and train this pre-evacuation behavior, especially for wildfires. 

Our current research initiates the conversation around packing for a wildfire and allows people to practice the packing scenario. We aim to address what items one should pack with them - sentimental (e.g., family photo album) and essential (e.g., medication)- with the intent of making participants more informed. 
VR can be uniquely effective for this type of training because it creates a safe and controlled scenario that allows users to go through the movements of picking up items and mentally deciding whether it's important to pack. Because of VR's novelty and immersive experiences, it can be a more engaging medium for practicing these steps than current training methods (i.e., reading information on a website). 
The following subsection describes the initial steps of designing and testing this application.

\subsection{Application}


Our VR experience is designed to address the packing aspect before a wildfire evacuation to help people who have not experienced a wildfire evacuation gain a sense of what that process would be like and a chance to practice \cite{crosby2024supporting}. This allows users to think through what to take, how long they would have to prepare, and where important items are located. This research stemmed from open-ended narrative interviews with people (n=9) who had experienced a wildfire \cite{loh2023toward}. These conversations demonstrated a need for improved pre-evacuation training. We also found a difference in mindset and confidence between interviewees with multiple evacuation experiences. This led us to our initial design and development of a VR evacuation game/application that allows users to understand the risk and level of preparation needed during an evacuation notice.

For this use case, the most important aspect of VR training is allowing players to situate themselves in the evacuation experience. They can experience the stress of packing under a time limit; they can see the glowing red haze outside their windows; they can go through the movements of running around a house trying to figure out what items may be important and what items they want to bring with them. In the game, players are placed in a house with the prompt that there is a wildfire in their area, and they should begin preparing for an evacuation notice. They are also given a finite amount of time, ten minutes, to walk through the home and pick up items. The VR experience allows physical movement around a house in a way a PC or mobile game could not. Players move by walking or using the joysticks and turning their heads to look around. This mechanic gives a more realistic replication of how they would look around their home to search for belongings to pack. At two minutes, there is a warning alarm, and the radio in the garage begins to change to an official evacuation notice stating that the player should leave as soon as possible. If the player evacuates in time, they are given a list of all the items they packed. If the player does not evacuate, they are only prompted with text that they did not make it out and to try playing again. 

We performed an initial pilot test (n=10) on the game and found overall positive experiences. All participants stated they would play the game again, and nine agreed that they felt like they were in a real evacuation. One participant wrote, ``The VR environment simulates the fire evacuation in real life very well. And also, making [the] decision to pick up limited stuff to carry from all the things you have in the house is a very educational way to teach me what is important when you actually encounter a fire.'' This pilot test provided a proof of concept of the evacuation game and informed the project's continued development. The next steps include providing a more immersive experience with things like sirens in the background, improving the controls, updating the tutorial, and potentially adding in-game character feedback (i.e., neighbors or roommates asking for help). We have also had several conversations with fire safety professionals who are interested in modifying our game for university dorm rooms that will engage with students on how to evacuate campus housing safely. 


In a follow-up study (n=20), we compared the VR experience to a mobile version with similar gameplay/structure and invited participants to reflect on and compare them. Participants commented on the more realistic and immersive nature of the VR experience, making it more likely to change their behavior. One participant noted, "Playing in virtual reality was much more immersive, and I think it probably was more effective in getting me to feel the experience of preparing to evacuate from a wildfire." Another participant said, "[The VR experience] looks much more realistic because of the possibility to move around the space more freely, and of course, because of the use of VR." This supports the idea that VR affordances can improve training potential for evacuation preparedness by creating an immersive and realistic experience. 

\section{Discussion}
 

While people may feel like they learned from, feel more prepared by, and were engaged with the game's training, it is important to check whether the game enacted real change. For example, did they go home and make a go-bag after playing the game? A longitudinal study should be performed with participants living in a wildfire-prone area before fire season (i.e., June) and then checking in with participants after fire season (i.e., November) to better understand the game's effectiveness. It is equally important to consistently check in with fire professionals to gauge the validity of this work and provide expert feedback.

Cybersickness is also a significantly persistent issue with VR technology \cite{davis2014systematic}. Within our pilot study, 4 participants reported feeling at least somewhat nauseous. When comparing the VR game to the mobile version in the follow-up study, we had 2 participants who could not finish the VR experience due to cybersickness. Additionally, when demoing this work at conferences, we've had several reports of extreme nausea. We could incorporate teleportation movement into the game, but teleportation may detract from one of the VR affordances we find most useful for this type of work. For the pilot study, players were only allowed to pick up two items at a time and had to go back and forth from the garage to pack items. The act of having to walk through the space virtually takes up more time and stresses the importance of exact and decisive decision-making. Another design possibility we tested was placing teleportation boxes around the house to pack items more quickly. However, when watching gameplay, players appeared to reach for whatever was nearby to pack rather than focusing on the necessity for each thing they grabbed. More discussion, playtesting, and development are needed to understand this specific gameplay mechanic better. 

The current game is made in Unity, and we use the Android build to install it on Meta's Quest 2 device, which runs an Android operating system. This approach makes the game self-contained on the headset, but lower-end hardware results in a reduced frame rate and may increase lag compared to running it on a computer and connecting the headset via a USB cable. The advantage is that cable-free means the player can physically stand and move around, though the trade-off of framerate and lag may increase cybersickness symptoms for those prone to motion sickness. Higher-end hardware such as the Valve Index or Apple Vision Pro would improve the virtual experience but become less accessible due to the high cost of those devices. Another option is to run the game on a high-end PC and stream it to the Meta Quest headset. However, latency could still be an issue depending on the streaming speed; it would also require a computer to run the experience, which may not be feasible when setting up demos at schools or other venues where it may be used. Ultimately, we find the standalone experience on the headset to offer the most accessible experience, so we focused our efforts on optimization techniques such as using low-poly models, reducing the number of objects in the scene with physics, and limiting particles and other effects. We also baked the lighting in the scene to reduce the impact of real-time lighting on the device.  

Beyond wildfire, we see the potential for such experiences to be applied to other natural hazards. As noted in the introduction, because of the effects of global warming, hurricanes are also expected to harm and affect human life on a greater scale. While our current VR game centers around a wildfire evacuation, much of the same pre-evacuation steps can be applied to those evacuating from a hurricane. Because of this, we can see this game being modified and applied to disseminating preparation on evacuation knowledge to an even greater audience.  

\section{Conclusion}

With the expected rise in the risk of wildfires and their impact on human life and property, it is increasingly important that people know how to prepare for an evacuation. We believe applying VR for pre-evacuation planning and packing can disseminate knowledge more effectively and engagingly than current training methods. VR allows for creating a safe and controlled environment to practice, provides an immersive experience with in-the-moment feedback, and enables the exploration of virtual environments with maneuvers similar to physical movements. 
To this end, our VR application has great potential to address the need for more informed and engaging pre-evacuation behavior training. 

\bibliographystyle{ACM-Reference-Format}
\bibliography{sample-base}


\end{document}